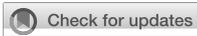

# Machine learning-based cloud resource allocation algorithms: a comprehensive comparative review

Deep Bodra[1]* and Sushil Khairnar[2]

[1]Information Systems Engineering and Management, Harrisburg University of Science and Technology, Harrisburg, PA, United States, [2]Virginia Tech, Blacksburg, VA, United States

Cloud resource allocation has emerged as a major challenge in modern computing environments, with organizations struggling to manage complex, dynamic workloads while optimizing performance and cost efficiency. Traditional heuristic approaches prove inadequate for handling the multi-objective optimization demands of existing cloud infrastructures. This paper presents a comparative analysis of state-of-the-art artificial intelligence and machine learning algorithms for resource allocation. We systematically evaluate 10 algorithms across four categories: Deep Reinforcement Learning approaches, Neural Network architectures, Traditional Machine Learning enhanced methods, and Multi-Agent systems. Analysis of published results demonstrates significant performance improvements across multiple metrics including makespan reduction, cost optimization, and energy efficiency gains compared to traditional methods. The findings reveal that hybrid architectures combining multiple artificial intelligence and machine learning techniques consistently outperform single-method approaches, with edge computing environments showing the highest deployment readiness. Our analysis provides critical insights for both academic researchers and industry practitioners seeking to implement next-generation cloud resource allocation strategies in increasingly complex and dynamic computing environments.

## 1 Introduction

### 1.1 Background and problem statement

Cloud computing has transformed the modern computing landscape with the global market reaching $912.77 billion in 2025 and projected to grow at a compound annual growth rate of 21.20% through 2034 (Precedence Research, 2025). This explosive growth reflects the critical role cloud infrastructure plays in supporting digital transformation initiatives across industries, as organizations increasingly rely on cloud services for scalability, flexibility, and cost optimization (Buyya et al., 2009; Armbrust et al., 2010). However, this rapid expansion has introduced challenges in resource allocation and management.

The complexity of cloud environments has grown exponentially as businesses adopt hybrid and multi-cloud strategies to meet operational requirements. Traditional resource allocation approaches based on heuristic algorithms and static provisioning models (Holland, 1992; Kennedy and Eberhart, 1995; Dorigo et al., 1996) have proven inadequate for handling the dynamic, heterogeneous, and multi-tenant nature of cloud infrastructures (Buyya et al., 2009;

Armbrust et al., 2010). These conventional methods struggle with the multi-objective optimization demands where performance, cost, energy efficiency, and quality of service must be optimized while adapting to fluctuating workloads and varying user demands.

Cloud resource management presents challenges for organizations with inefficient resource allocation leading to substantial waste and increased operational costs. There is a need for more sophisticated, adaptive resource allocation mechanisms that can intelligently respond to dynamic cloud environments.

Traditional approaches to cloud resource allocation, including First-Fit, Best-Fit, and basic optimization algorithms (Zhang et al., 2010; Beloglazov and Buyya, 2012) face fundamental limitations when confronted with the scale and complexity of modern cloud deployments. These methods typically rely on predefined rules and static policies that cannot adapt to changing workload patterns, user behaviors, or infrastructure conditions. As cloud environments continue to scale with exponential data growth, the limitations of conventional approaches come into effect.

## 1.2 Research contributions

This paper addresses these challenges by presenting a comparative analysis of state-of-the-art artificial intelligence and machine learning algorithms specifically designed for cloud resource allocation. This paper addresses the identified gaps in current survey literature by presenting a focused analysis of state-of-the-art artificial intelligence and machine learning algorithms specifically designed for cloud resource allocation, with emphasis on recent developments from 2022 to 2024. While existing comprehensive surveys have established foundational coverage of ML-centric resource management and provided systematic analysis of resource allocation strategies, our work contributes to the field through several distinct approaches.

We provide a systematic in-depth analysis of 10 cutting-edge AI/ML algorithms published in recent high-impact venues, covering Deep Reinforcement Learning approaches (PPO for D2D-assisted MEC, ATSIA3C, Rainbow DQN), Neural Network architectures (DPSO-GA, VSBG, BiGRU with DWT), Traditional ML enhanced methods (enhanced-Kernel SVM, N2TC-GATA), and multi-agent systems (multi-agent DRL for container allocation, Industrial Federated DDPG).

Our work extends beyond algorithmic classification by providing critical analysis of implementation trade-offs, convergence properties, and scalability limitations that affect real-world deployment decisions. We synthesize practical insights for industry adoption, examining the performance characteristics of algorithms that integrate multiple AI/ML techniques compared to single-method approaches. The analysis identifies key trends in hybrid architecture design and federated learning integration that distinguish recent developments from earlier ML-based resource allocation research.

The practical significance of this research builds upon the growing integration of artificial intelligence with cloud computing infrastructures (Sutton and Barto, 2018; Mnih et al., 2015; Silver et al., 2016). As organizations increasingly deploy AI and machine learning services, the demand for intelligent resource allocation mechanisms capable of efficiently supporting these workloads becomes critical. Our analysis provides both academic researchers and industry practitioners with focused insights for implementing next-generation cloud resource allocation strategies that leverage the most recent AI/ML innovations.

## 2 Related work and background

### 2.1 Cloud resource allocation fundamentals

Cloud resource allocation requires the systematic assignment of computational resources including CPU, memory, storage, and network bandwidth among competing user requests to optimize system performance while maintaining service level agreements. The fundamental challenge lies in efficiently mapping heterogeneous user workloads to distributed physical resources while satisfying multiple conflicting objectives such as minimizing execution time, reducing energy consumption, and maximizing resource utilization.

Traditional resource allocation approaches in cloud computing environments primarily rely on heuristic algorithms, meta-heuristic techniques, and hybrid methods. Heuristic algorithms, including First-Fit, Best-Fit, and Greedy algorithms (Zhang et al., 2010), provide intuitive solutions based on empirical construction with lower computational complexity and predictable worst-case performance. These methods typically use simple rules such as selecting the first available resource that meets minimum requirements or choosing resources with the smallest remaining capacity after allocation.

Meta-heuristic approaches, including Genetic Algorithm (GA) (Holland, 1992), Particle Swarm Optimization (PSO) (Kennedy and Eberhart, 1995), and Ant Colony Optimization (ACO) (Dorigo et al., 1996), have gained prominence for addressing the NP-hard nature of resource allocation problems. These algorithms employ population-based search mechanisms to explore solution spaces more comprehensively than heuristic methods, often achieving superior optimization results at the cost of increased computational complexity. Hybrid approaches combine multiple optimization techniques, leveraging the strengths of different algorithms to address specific aspects of the resource allocation problem.

Performance evaluation in cloud resource allocation relies on Quality of Service (QoS) metrics (Garg et al., 2013; Zhang et al., 2010) that capture various aspects of system behavior and user experience. Critical performance indicators include response time, throughput, resource utilization, availability, and cost efficiency. Response time measures the latency between request submission and completion, while throughput quantifies the system's capacity to process requests within specific time periods. Resource utilization metrics assess the efficiency of hardware usage, preventing both over-provisioning and under-utilization scenarios that lead to economic inefficiencies (Beloglazov and Buyya, 2012; Li et al., 2013).

Advanced performance evaluation frameworks incorporate multi-dimensional metrics that address the complexity of modern cloud environments. These include scalability measures that evaluate system behavior under varying loads, reliability indicators that assess fault tolerance capabilities, and energy efficiency metrics that quantify power consumption relative to computational output. The integration of Service Level Objectives (SLO) and Service Level Agreements (SLA) (Buyya et al., 2009) provides contractual frameworks for performance measurement, establishing measurable targets for system behavior.

## 2.2 AI/ML evolution in resource management

The paradigm shift from reactive to predictive resource allocation represents a fundamental transformation in cloud computing management strategies. Traditional reactive approaches respond to resource demands after they occur, leading to suboptimal performance during peak loads and resource waste during low-demand periods. In contrast, AI/ML-enabled predictive allocation systems analyze historical patterns, workload characteristics, and system behaviors to anticipate future resource requirements, enabling proactive resource provisioning and optimization.

Machine learning techniques have demonstrated significant potential in addressing the dynamic and complex nature of cloud resource allocation challenges. Supervised learning approaches utilize historical workload data to predict future resource demands, while unsupervised learning methods identify hidden patterns in resource usage that can inform allocation strategies. Reinforcement learning techniques enable adaptive resource allocation through continuous interaction with the cloud environment, learning optimal policies through trial-and-error experiences (Sutton and Barto, 2018).

Deep learning approaches, particularly Deep Reinforcement Learning (DRL) (Sutton and Barto, 2018; Mnih et al., 2015; Silver et al., 2016), have emerged as powerful solutions for complex resource allocation scenarios where traditional algorithms struggle. DRL combines the pattern recognition capabilities of deep neural networks with the decision-making strengths of reinforcement learning, enabling systems to handle high-dimensional state spaces and complex optimization objectives. These approaches have shown particular effectiveness in scenarios with dynamic workloads, heterogeneous resources, and multi-objective optimization requirements.

Existing survey literature reveals significant gaps in comparative analysis of recent AI/ML approaches for cloud resource allocation. While several surveys have addressed specific aspects such as energy efficiency, load balancing techniques, or particular algorithm categories, there remains a lack of systematic evaluation that encompasses the breadth of AI/ML techniques currently being developed (Zhang et al., 2010; Sutton and Barto, 2018). Most existing reviews focus on algorithmic classifications rather than quantitative performance comparisons, limiting their utility for practical implementation decisions.

Furthermore, the rapid evolution of AI/ML techniques has outpaced survey efforts, with many cutting-edge algorithms remaining unanalyzed in existing literature (Lu et al., 2017). The integration of modern deep learning architectures, hybrid optimization approaches, and multi-agent systems represents an emerging research that requires systematic investigation. Our work addresses these gaps by providing an analysis of state-of-the-art AI/ML algorithms specifically designed for cloud resource allocation, offering both technical insights and practical implementation guidance for the next generation of intelligent cloud management systems. See Table 1 for an overview.

The table presents improvements to provide a view of algorithmic capabilities. The diversity of metrics across studies reflects different optimization priorities: healthcare applications prioritize latency reduction, mobile systems balance energy and performance, industrial environments focus on energy efficiency, and enterprise systems optimize for cost and resource utilization. Direct cross-metric comparison is limited by heterogeneous evaluation environments, but the multi-metric view illustrates algorithmic trade-offs and suitability for different deployment scenarios.

## 3 Methodology

This systematic review employs a structured approach adapted from PRISMA guidelines to ensure transparency and reproducibility in literature selection and analysis. The literature search strategy encompassed multiple academic databases including IEEE Xplore Digital Library, ACM Digital Library, ScienceDirect, Springer Link, and arXiv preprint server, focusing on publications from January 2022 to December 2024 to capture recent advances in AI/ML-based cloud resource allocation. The search employed Boolean operators combining "cloud resource allocation" OR "cloud resource management" with "machine learning" OR "artificial intelligence" OR "deep learning" and category-specific terms including "DQN," "policy gradient," "CNN," "LSTM," "SVM," "genetic algorithm," and "federated learning."

The selection process followed rigorous inclusion and exclusion criteria targeting algorithms specifically designed for cloud computing resource allocation with novel AI/ML approaches, quantitative performance evaluation, and practical implementation considerations from 2022 to 2024 publications. The systematic selection proceeded through three stages: initial screening of 2,847 papers through title and abstract evaluation resulted in 287 papers for full-text review; detailed assessment of technical content, algorithmic novelty, and experimental validation yielded 45 papers meeting all criteria; final selection focused on 10 algorithms representing methodological diversity across four categories, performance excellence exceeding 10% improvement in key metrics, implementation readiness, experimental rigor, and preference for recent 2024 publications representing cutting-edge research.

The analysis framework addresses the fundamental challenge of comparing algorithms evaluated under different conditions and methodologies. Given the heterogeneous nature of evaluation environments across selected papers, the comparative analysis focuses on relative improvement percentages rather than absolute values, categorizes algorithms by problem type and evaluation context, and provides qualitative assessment where quantitative comparison proves infeasible. Performance metrics normalization emphasizes primary metrics including execution time improvement, cost optimization, and energy efficiency, alongside secondary metrics such as resource utilization, QoS satisfaction, and scalability measures. Technical innovation assessment examines algorithmic architecture including neural network design and optimization techniques, problem formulation encompassing state space representation and objective functions, and learning mechanisms covering training procedures and convergence properties.

The methodology acknowledges significant limitations inherent in cross-study comparison of machine learning algorithms for cloud resource allocation. Evaluation environment diversity across different simulators such as CloudSim and iFogSim, varying datasets including Google cluster traces and Alibaba traces, and different experimental setups limit direct quantitative comparison between algorithms. Baseline variation occurs as different papers employ different baseline algorithms for comparison, while metric heterogeneity reflects varying performance metrics and measurement methodologies across studies. Scale differences manifest through evaluation conducted at different scales regarding number of virtual machines, tasks, and time periods. These limitations necessitate careful interpretation of comparative results and emphasize the importance of relative performance improvements rather than absolute metric values in evaluating algorithmic effectiveness.

TABLE 1 Overview of analyzed machine learning algorithms for cloud resource allocation.

| Category | Algorithm | Key innovation | Primary application | Improvements |
|---|---|---|---|---|
| Deep reinforcement learning | PPO for D2D-MEC | Policy gradient with clipping | Mobile edge computing | 35–45% execution time improvement, 40–50% energy savings, 30–40% cost reduction |
| | ATSIA3C | Residual CNN + A3C | Multi-cloud task scheduling | 70.49% makespan reduction, 77.42% cost optimization, 74.24% energy improvement |
| | Rainbow DQN | Six DQN enhancements | IoT edge-cloud systems | 43.1% utility enhancement, 29.8% energy efficiency, 27.5% latency reduction |
| Neural network architecture | DPSO-GA | CNN-LSTM + meta-heuristic | Cloud load balancing | 13.3% cost reduction, energy optimization, prediction MAE improvements |
| | VSBG | VMD + Hybrid LSTM | Workload prediction | RMSLE 0.03 vs. 0.89 baseline, MSE Reduction, $R^2 = 0.97$ |
| | DWT-BiGRU | Wavelet + attention | Host load prediction | 15.4% MAPE (high volatility), RMSE reduction across machine types |
| Traditional ML enhanced | Enhanced-Kernel SVM | Novel kernel fusion | Healthcare fog-cloud | 73.88% execution time improvement, 90% latency reduction vs. cloud-only |
| | N2TC-GATA | Neural classification + GA | Multi-objective allocation | 13.3% cost, 12.1% response time, 3.2% execution time improvement |
| Multi-agent based | Multi-agent DRL | Cooperative-competitive agents | Container allocation | 28% overall runtime improvement, enhanced container placement efficiency |
| | IF-DDPG | Federated DDPG | Industrial edge computing | 50.5% energy consumption reduction, 15.2–31.75 improvement vs. other DRL methods |

# 4 Algorithm analysis

This section presents a comprehensive technical analysis of 10 state-of-the-art machine learning algorithms for cloud resource allocation, organized into four distinct categories based on their underlying methodological approaches. Each category represents a different paradigm in applying machine learning techniques to address the complex challenges of resource allocation in modern cloud computing environments. The algorithms analyzed span recent developments from leading research venues and demonstrate significant advances over traditional heuristic approaches. Figure 1 presents the taxonomic organization of the 10 analyzed algorithms across four main methodological categories, providing a structural overview of the algorithmic landscape examined in this survey.

## 4.1 Deep reinforcement learning approaches

Deep reinforcement learning has emerged as a paradigm for cloud resource allocation combining the decision-making capabilities of reinforcement learning with the pattern recognition capability of deep neural networks (Sutton and Barto, 2018; Mnih et al., 2015; Mao et al., 2017). These approaches excel in dynamic environments where traditional optimization methods struggle with the complexity of multi-dimensional state spaces, temporal dependencies, and continuous adaptation requirements. The algorithms in this category demonstrate significant advances in handling uncertainty, learning from experience, and making optimal allocation decisions in real-time cloud environments.

The three algorithms examined—Proximal Policy Optimization for Device-to-Device-assisted Mobile Edge Computing, Adaptive Task Scheduler using Improved Asynchronous Advantage Actor-Critic, and Rainbow Deep Q-Network for edge-cloud systems—represent different aspects of deep reinforcement learning innovation. They showcase advances in policy optimization, actor-critic architectures, and value-based learning, respectively, while addressing specific challenges in mobile edge computing, multi-cloud task scheduling, and hierarchical resource allocation.

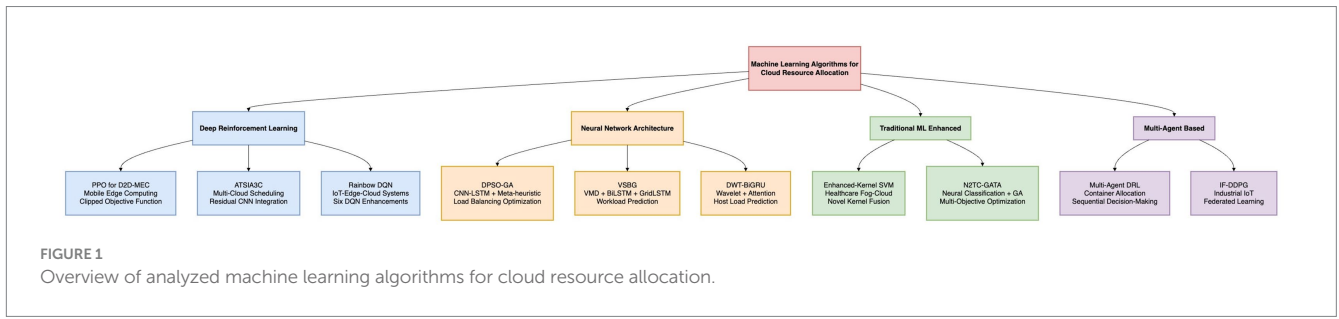

FIGURE 1
Overview of analyzed machine learning algorithms for cloud resource allocation.

### 4.1.1 Proximal policy optimization for D2D-assisted MEC

Proximal Policy Optimization (PPO) has been successfully adapted for Mobile Edge Computing environments through its clipped objective function mechanism that prevents destructive policy updates while maintaining stable learning characteristics (Cao et al., 2024). The algorithm employs both policy and value networks, where the policy network outputs probability distributions over offloading actions, and the value network estimates state values for advantage calculation using Generalized Advantage Estimation.

PPO addresses the fundamental challenge of intelligent task offloading decisions across three-tier architectures comprising mobile devices, edge servers, and cloud data centers. The approach formulates resource allocation as a Markov Decision Process that simultaneously optimizes execution time, energy consumption, and monetary costs while enabling devices with limited computational capabilities to make critical decisions about task execution location.

PPO's technical innovation lies in its clipped objective function mechanism (Schulman et al., 2017; Sutton and Barto, 2018) that prevents destructive policy updates while maintaining stable learning characteristics. The algorithm employs both policy and value networks, where the policy network outputs probability distributions over offloading actions, and the value network estimates state values for advantage calculation. The clipped surrogate objective function $J(\theta)$:

$$J(\theta) = \widehat{E}_t \left[ \min\left(r_t(\theta)\widehat{A}_t, \mathrm{clip}\left(r_t(\theta), 1-\varepsilon, 1+\varepsilon\right)\widehat{A}_t\right) \right]$$

ensures policy updates remain within safe bounds, preventing instability issues common in traditional policy gradient methods.

Experimental evaluation demonstrates significant performance improvements across multiple metrics compared to baseline algorithms. The PPO approach achieves 35–45% improvement in execution time reduction versus random offloading, 40–50% energy savings compared to local execution, and 30–40% cost reduction versus always-cloud strategies (Cao et al., 2024). The algorithm shows convergence within 500–1000 episodes and requires 20–30% fewer samples than vanilla policy gradient methods, demonstrating superior sample efficiency essential for practical deployment scenarios.

The algorithm requires specialized infrastructure for policy network training, GPU acceleration for neural network processing, and integration with mobile edge computing simulation environments. Implementation complexity involves hyperparameter tuning for clipped objective functions, state space design for multi-tier environments, and reward function formulation for multi-objective optimization scenarios.

PPO represents a significant advancement in mobile edge computing resource allocation by enabling intelligent, adaptive task offloading that responds to dynamic network conditions, device constraints, and application requirements. The approach demonstrates how policy gradient methods can be stabilized and applied to real-time resource allocation scenarios with multiple conflicting objectives.

### 4.1.2 Adaptive task scheduler using improved A3C (ATSIA3C)

The Adaptive Task Scheduler using Improved Asynchronous Advantage Actor-Critic introduces architectural innovation that replaces standard A3C fully connected networks with Residual Convolutional Neural Networks (Hussain et al., 2024). This modification enables more effective learning in complex multi-cloud environments by preserving gradient flow and improving feature extraction capabilities through skip connections that mitigate vanishing gradient problems.

ATSIA3C directly addresses the fundamental challenge of mapping heterogeneous tasks to distributed cloud resources where traditional algorithms struggle with varying task lengths, runtime capacities, and diverse resource requirements (Mnih et al., 2016; Calheiros et al., 2011). The algorithm incorporates intelligent task segmentation that analyzes incoming tasks and partitions them into subtasks based on computational complexity, memory requirements, communication dependencies, and execution time constraints.

The task segmentation mechanism represents a critical innovation that analyzes incoming tasks and partitions them into subtasks based on computational complexity, memory requirements, communication dependencies, and execution time constraints. Large monolithic tasks undergo intelligent decomposition that considers inter-task dependencies, data locality requirements, and communication overhead minimization. The segmentation algorithm employs graph-based analysis to identify task components that can execute independently while maintaining data consistency and dependency satisfaction.

The algorithm requires CloudSim simulation environment familiarity, multi-cloud API integration capabilities, and distributed computing infrastructure for training residual CNN components. Implementation complexity involves residual CNN configuration, A3C hyperparameter tuning, task dependency graph construction, and multi-cloud resource monitoring systems.

Evaluation using CloudSim toolkit demonstrates exceptional improvements: 70.49% reduction in makespan compared to baseline algorithms including RATS-HM, MOABCQ, and AINN-BPSO, 77.42% improvement in resource cost optimization on average, and 74.24% improvement in energy consumption optimization (Hussain et al., 2024). These results indicate substantial advancement in overall task completion time while achieving cost-efficient resource selection across multi-cloud environments.

ATSIA3C represents the evolution from reactive to predictive multi-cloud scheduling, demonstrating how architectural innovations in deep reinforcement learning can address real-world resource allocation complexity. The approach establishes new performance benchmarks for multi-cloud task scheduling while maintaining practical deployment feasibility for enterprise environments.

### 4.1.3 Rainbow DQN for edge-cloud systems

Rainbow Deep Q-Network integrates six key enhancements to standard DQN algorithms: Double Q-Learning, Prioritized Experience Replay, Dueling Networks, Multi-step Learning, Distributional Reinforcement Learning, and Noisy Networks (Aloqaily et al., 2024). Each component addresses specific limitations of traditional DQN approaches through comprehensive enhancement strategies that improve sample efficiency, reduce overestimation bias, and enable more sophisticated exploration.

The algorithm operates within three-layer Device-to-Device-Edge-Cloud architectures, addressing the complexity of hierarchical resource allocation across IoT devices, edge nodes, and cloud infrastructure. Rainbow DQN handles the fundamental challenge of intelligent resource allocation decisions across multiple tiers with varying computational capabilities, communication constraints, and latency requirements.

Experimental evaluation demonstrates significant performance improvements across all key metrics in realistic IoT deployment scenarios. Rainbow DQN achieves 29.8% improvement in energy efficiency, 27.5% reduction in latency, and 43.1% increase in utility compared to Double DQN baselines when tested with 100 IoT devices (Aloqaily et al., 2024). Performance gains scale effectively with network size, showing 32% energy efficiency improvement and 50% utility enhancement with 300 devices.

The algorithm requires specialized deep learning infrastructure supporting six integrated enhancement components, extensive memory for prioritized experience replay buffers, and distributed computing capabilities for multi-step learning processes. Implementation involves complex hyperparameter coordination across multiple enhancement mechanisms and sophisticated exploration strategy management.

Rainbow DQN advances the field by demonstrating how comprehensive integration of DQN enhancements can address the complexity of modern distributed computing environments (Hessel et al., 2018). The approach provides a foundation for hierarchical resource allocation in IoT-edge-cloud scenarios while establishing new performance standards for value-based reinforcement learning in resource management.

## 4.2 Neural network architecture based approaches

Neural network architectures represent a fundamental paradigm in applying machine learning to cloud resource allocation, leveraging deep learning's pattern recognition capabilities to predict workload demands and optimize resource distribution. These approaches excel in extracting complex non-linear relationships from historical data, enabling proactive resource allocation that anticipates future demands rather than merely reacting to current conditions. The three algorithms examined—DPSO-GA hybrid optimization, VSBG multi-modal prediction, and BiGRU with discrete wavelet transformation—represent different aspects of neural network innovation in cloud computing. They showcase advances in hybrid optimization techniques, signal processing integration, and bidirectional temporal modeling, respectively, while addressing specific challenges in load balancing, workload prediction, and host utilization forecasting.

### 4.2.1 Deep learning with particle swarm intelligence and genetic algorithm (DPSO-GA)

The DPSO-GA algorithm introduces a hybrid approach that combines Convolutional Neural Networks (He et al., 2016) and Long Short-Term Memory networks (Hochreiter and Schmidhuber, 1997) with Particle Swarm Optimization (Kennedy and Eberhart, 1995) and Genetic Algorithm optimization (Awad et al., 2024; Holland, 1992). The innovation lies in a two-phase optimization strategy where PSO-GA optimization performs intelligent hyperparameter selection while CNN-LSTM neural networks provide sophisticated workload prediction and resource allocation decisions.

DPSO-GA addresses critical challenges in cloud environments where dynamic workload patterns create resource imbalances leading to over-provisioning, under-utilization, and inefficient energy consumption. The algorithm provides intelligent workload prediction and proactive resource allocation that anticipates demand changes and optimizes resource distribution before performance degradation occurs.

Experimental evaluation using Google cluster workload traces (Reiss et al., 2011; Verma et al., 2015) and CloudSim simulation (Calheiros et al., 2011) demonstrates significant performance improvements across multiple metrics. The DPSO-GA approach achieves optimal waiting times of 10.2 s for 10 VM configurations, with energy consumption ranging from 201.77 KWh for 50 tasks to 809.91 KWh for 1500 tasks (Awad et al., 2024). Multi-variate analysis shows MAE improvements in storage prediction (0.18 vs. 0.25), processing power (0.29 vs. 0.37), and memory utilization (0.024 vs. 0.036).

The algorithm requires deep learning framework infrastructure supporting CNN-LSTM training, meta-heuristic optimization libraries for PSO-GA implementation, and CloudSim simulation environment capabilities. Implementation complexity involves hyperparameter optimization coordination, neural network architecture design, and integration of multiple optimization paradigms.

DPSO-GA demonstrates the effectiveness of hybrid optimization approaches that combine deep learning pattern recognition with meta-heuristic global optimization. The algorithm establishes new paradigms for proactive cloud resource management while providing practical frameworks for organizations seeking to implement intelligent load balancing systems.

### 4.2.2 Variational mode decomposition with bidirectional and grid LSTM (VSBG)

The VSBG algorithm introduces systematic combination of advanced signal processing techniques with hybrid deep learning architectures through integration of Variational Mode Decomposition,

Savitzky–Golay filtering, Bidirectional LSTM (Hochreiter and Schmidhuber, 1997), and Grid LSTM components (Yuan et al., 2024). The innovation enables superior prediction accuracy by decomposing nonstationary workload time series into manageable components that can be individually processed and optimized for maximum predictive performance.

VSBG addresses fundamental challenges in cloud workload prediction where traditional approaches fail to handle complex, multi-scale temporal patterns characteristic of modern cloud environments. The algorithm processes workloads that exhibit nonstationary behavior with multiple overlapping periodicities, sudden spikes, gradual trends, and random fluctuations that create prediction difficulties for conventional methods.

Performance evaluation using heterogeneous datasets from Google cluster traces (672,003 jobs over 29 days) and Alibaba cluster traces (4M jobs over 8 days) demonstrates exceptional accuracy improvements. For Google cluster workload prediction, VSBG achieves RMSLE of 0.03 compared to 0.89 for ARIMA and 0.81 for LSTM (Yuan et al., 2024), with MSE of 7583 ± 4 versus 18,133 ± 4 for ARIMA and 10,942 ± 2 for LSTM. The $R^2$ coefficient reaches 0.97 compared to 0.81 for ARIMA and 0.89 for LSTM.

The algorithm requires advanced signal processing libraries supporting VMD implementation, specialized LSTM frameworks capable of bidirectional and grid architectures, and substantial computational resources for multi-modal prediction processing. Implementation involves VMD parameter tuning, hybrid LSTM configuration, and integration of multiple signal processing and deep learning components.

VSBG represents a significant advancement in cloud workload prediction by systematically addressing signal processing challenges that limit traditional approaches. The algorithm enables more accurate resource allocation decisions and improved quality of service guarantees through predictive rather than reactive resource management strategies.

### 4.2.3 Discrete wavelet transformation with bidirectional GRU (DWT-BiGRU)

The DWT-BiGRU algorithm combines Discrete Wavelet Transformation, Bidirectional Gated Recurrent Units (Cho et al., 2014), and attention mechanisms (Vaswani et al., 2017) through a three-stage hybrid approach specifically targeting CPU utilization prediction (Almalki et al., 2022). The innovation employs Mallat's algorithm for three-level signal decomposition, separating nonstationary host load traces into multiple frequency components that enable targeted processing and improved prediction accuracy.

DWT-BiGRU addresses critical challenges in host load prediction where accurate forecasting of CPU utilization patterns directly impacts resource allocation decisions, auto-scaling mechanisms, and quality of service guarantees. The algorithm handles nonstationary cloud workloads that exhibit multiple time scales, sudden changes, and complex patterns influenced by user behavior, application characteristics, and system dynamics.

Experimental validation using Google cluster (672,074 jobs over 29 days) and Alibaba cluster (1300 + machines over 12 h) datasets demonstrates significant performance improvements across diverse machine configurations. For Google cluster Machine G1 with low average CPU and high volatility, DWT-BiGRU-attention achieves MAPE of 15.4% compared to 34.2% for SVR and 23.1% for LSTM (Almalki et al., 2022), with RMSE of 3.1 versus 10.2 and 6.5, respectively.

The algorithm requires signal processing libraries supporting discrete wavelet transformation, specialized GRU implementations with bidirectional capabilities, and attention mechanism frameworks for encoder–decoder architectures. Implementation involves wavelet parameter selection, BiGRU network configuration, attention mechanism tuning, and integration of signal processing with deep learning components.

DWT-BiGRU advances the field by demonstrating how signal processing techniques can be systematically integrated with recurrent neural networks to address prediction challenges in dynamic cloud environments. The approach provides practical frameworks for organizations seeking to implement intelligent host load prediction for improved resource allocation decisions.

## 4.3 Traditional machine learning enhanced approaches

### 4.3.1 Service-aware hierarchical fog-cloud resource mapping with enhanced-kernel SVM

The Enhanced-Kernel SVM algorithm introduces novel kernel design that combines multiple kernel functions with traditional Support Vector Machine classification (Cortes and Vapnik, 1995; Cristianini and Shawe-Taylor, 2000) to achieve superior classification accuracy for healthcare-specific task categorization (AlZailaa et al., 2024). The innovation integrates cross-correlation for measuring task similarity through symmetric features, convolution operations for enhanced accuracy through reversed similarity analysis, and auto-correlation for capturing self-similarity within task classes.

Enhanced-Kernel SVM addresses the critical challenge of latency-sensitive healthcare applications where real-time processing capabilities and quality of service guarantees are essential for patient safety and care effectiveness (Bonomi et al., 2012). The algorithm operates in complex healthcare computing environments requiring ultra-low latency for critical medical monitoring, high reliability for life-safety systems, and strict privacy protection for sensitive medical data (Shi et al., 2016).

The algorithm requires iFogSim simulation environment capabilities, specialized SVM libraries supporting custom kernel design, and healthcare domain expertise for task classification and priority determination. Implementation involves kernel fusion parameter tuning, healthcare-specific feature extraction, and integration with fog-cloud infrastructure management systems.

Enhanced-Kernel SVM establishes new paradigms for domain-specific resource allocation by demonstrating how traditional machine learning techniques can be enhanced through specialized kernel design. The approach provides practical frameworks for healthcare organizations seeking to implement intelligent fog-cloud resource allocation with strict latency and reliability requirements.

Experimental evaluation using iFogSim simulator (Gupta et al., 2017) demonstrates exceptional performance improvements across multiple healthcare scenarios. The Enhanced-Kernel SVM achieves 73.88% improvement in execution time for critical tasks (0.23ms versus 0.92ms for baseline FCFS algorithm) and 52.01% improvement for non-critical tasks (AlZailaa et al., 2024). Latency analysis shows remarkable 90% reduction compared to cloud-only architectures, with fog-cloud hybrid deployment achieving 11.4 ms to 23.06 ms latency for 20–60 sensors versus 88.14 ms to 331.81 ms for cloud-only configurations.

### 4.3.2 Neural network task classification with genetic algorithm task assignment (N2TC-GATA)

The N2TC-GATA algorithm combines Neural Network Task Classification with Genetic Algorithm Task (Holland, 1992) Assignment through a two-stage methodology that leverages pattern recognition capabilities for intelligent task classification with global optimization strengths for optimal resource assignment (Manaseer and Ali, 2024). The innovation employs a feed-forward back-propagation neural network with 20 hidden layers specifically designed for task priority classification integrated with genetic algorithm optimization using decimal chromosome encoding.

N2TC-GATA addresses optimization challenges in cloud environments where diverse task characteristics, heterogeneous resource capabilities, and multiple conflicting objectives create scheduling scenarios that exceed the capabilities of traditional single-method approaches. The algorithm handles complex scenarios requiring simultaneous optimization of execution time, cost considerations, and system efficiency factors.

The algorithm requires neural network frameworks supporting multi-layer feed-forward architectures, genetic algorithm optimization libraries with decimal encoding capabilities, and substantial computational resources for population-based optimization with 500 chromosomes. Implementation involves neural network training procedures, genetic algorithm parameter tuning, and integration of classification and optimization components.

N2TC-GATA demonstrates the effectiveness of combining neural network classification with evolutionary optimization for multi-objective cloud resource allocation. The approach provides practical frameworks for organizations seeking balanced optimization across multiple performance criteria while maintaining computational efficiency and implementation feasibility.

Performance evaluation using Google cluster-traces v3 dataset with 405,894 task records demonstrates significant improvements across multiple metrics. The N2TC-GATA approach achieves 3.2% reduction in execution time, 13.3% improvement in cost efficiency, and 12.1% enhancement in response time (Manaseer and Ali, 2024). Neural network training demonstrates efficient convergence within 27 epochs, while genetic algorithm optimization shows consistent fitness improvement across generations.

## 4.4 Multi-agent based approaches

Multi-agent based approaches represent a paradigm shift toward distributed intelligence (Stone and Veloso, 2000; Tampuu et al., 2017; Dragoni et al., 2017) in cloud resource allocation, leveraging the collective decision-making capabilities of multiple autonomous agents to address complex optimization challenges that exceed the capabilities of centralized algorithms. These approaches excel in scenarios requiring coordinated resource allocation across heterogeneous environments, dynamic adaptation to changing conditions, and scalable solutions that can accommodate growing system complexity.

The two algorithms examined—Multi-Agent Deep Reinforcement Learning for container allocation and Industrial Federated Deep Deterministic Policy Gradient (IF-DDPG) for edge computing—represent different aspects of multi-agent innovation. They showcase advances in coordinated container placement optimization and federated energy-efficient resource allocation, respectively, while addressing specific challenges in cloud orchestration systems and industrial Internet of Things (IoT) environments.

### 4.4.1 Container allocation using multi-agent deep reinforcement learning

The Multi-Agent Deep Reinforcement Learning framework employs multiple autonomous agents operating within a mixed cooperative-competitive environment that optimizes both individual container performance and overall system efficiency (Chen et al., 2024). The innovation implements sequential decision-making where agents operate in predetermined serial order, ensuring coordinated placement decisions while maintaining individual agent autonomy through comprehensive system state observation including container requirements, previous agents' decisions, and real-time server utilization metrics (Burns et al., 2019).

Multi-Agent DRL addresses complex container orchestration challenges in modern cloud environments where microservices architectures, containerized applications, and dynamic scaling requirements create resource allocation scenarios that exceed the capabilities of traditional scheduling approaches. The algorithm handles complex decisions about resource assignment, inter-container communication optimization, load balancing, and quality of service maintenance across heterogeneous infrastructure.

The algorithm requires distributed computing infrastructure supporting multi-agent coordination, specialized deep reinforcement learning frameworks capable of sequential decision-making, and container orchestration platform integration for real-world deployment. Implementation involves agent coordination protocol design, LSTM network configuration for individual agents, and integration with existing container management systems.

Multi-Agent DRL establishes new paradigms for distributed intelligent resource allocation by demonstrating how multiple autonomous agents can coordinate effectively to achieve superior system-wide performance. The approach provides practical frameworks for organizations seeking to implement intelligent container orchestration with improved performance over traditional scheduling algorithms.

The multi-agent coordination mechanism employs sequential decision-making where agents operate in predetermined serial order, ensuring coordinated placement decisions while maintaining individual agent autonomy. Each agent observes comprehensive system information including current container requirements, previous agents' decisions, and real-time server utilization metrics. The state space formulation of agent $i$ is given by

$$S_i(t) = \begin{cases} \{q_i(t)\}, \{a_1(t), \ldots, a_{i-1}(t), a_{i+1}(t-1), \ldots, a_K \\ (t-1)\}, \{u_1(t-1), \ldots, u_M(t-1)\} \end{cases}$$

where $q_i(t)$ denotes the container for which the $i$th agent is responsible at decision round $t$ the $i$th agent's task requirement at round $t$. The previous decisions made by the agents that preceded it in the current round of decisions are denoted by the action vector $a_j(t)$. $u_j(t)$ stands for the utilization of the servers concerning the previous decision round. This incorporates both local container specifications and global system state information, enabling informed decision-making that considers system-wide impact while maintaining agent autonomy.

Experimental evaluation in real private cloud environments with eight heterogeneous servers demonstrates exceptional performance improvements. The MADRL approach achieves 28% overall runtime improvement compared to existing techniques, with LSTM-based agents showing 25–32% superiority over Kubernetes, Best-Fit, and Max-Fit algorithms (Chen et al., 2024). Performance advantages are particularly pronounced for communication-intensive container batches, where intelligent placement of communicating containers yields significant efficiency gains through reduced network overhead and improved data locality.

### 4.4.2 Industrial federated deep deterministic policy gradient (IF-DDPG)

The Industrial Federated Deep Deterministic Policy Gradient algorithm integrates federated learning (McMahan et al., 2017; Li et al., 2020) with deep reinforcement learning for energy-efficient edge computing offloading and resource allocation in Industrial Internet environments (Kumar et al., 2024). The innovation combines Deep Deterministic Policy Gradient reinforcement learning with federated learning architecture to achieve collaborative optimization without compromising data privacy essential for industrial competitiveness through sophisticated parameter aggregation mechanisms (GDPR.eu, 2018).

IF-DDPG addresses critical challenges of optimizing task offloading decisions, communication resource allocation, and computing resource allocation across multiple edge servers and mobile industrial terminal devices while maintaining strict data privacy requirements (Dwork, 2008) essential for industrial applications. The algorithm handles industrial environments where manufacturing processes, supply chain management, quality control systems, and equipment monitoring create diverse computational workloads requiring intelligent resource allocation.

The algorithm requires federated learning infrastructure supporting distributed model training, specialized DDPG implementation capable of industrial IoT integration, and edge computing platforms with communication and computation resource management capabilities. Implementation involves federated parameter aggregation protocols, local DDPG training procedures, and integration with industrial IoT device management systems.

IF-DDPG advances the field by demonstrating how federated learning can be systematically integrated with deep reinforcement learning to address privacy-preserving collaborative optimization in industrial environments. The approach establishes new standards for energy-efficient resource allocation while providing practical frameworks for industrial organizations seeking intelligent edge computing optimization.

Experimental evaluation in realistic industrial environments with four edge servers and four terminal devices across $100 \times 100$ m$^2$ factory areas demonstrate significant energy consumption reductions compared to baseline algorithms. IF-DDPG achieves 15.2% energy reduction versus traditional DDPG, 31.7% improvement over Deep Double Q-Network (DDQN), 38.7% enhancement compared to Deep Q-Network (DQN), and remarkable 50.5% reduction versus Actor-Critic (AC) algorithms (Kumar et al., 2024). Convergence analysis shows faster, and more stable learning compared to baseline approaches, with optimal learning rates providing superior performance characteristics.

## 5 Evaluation methodology and cross-study comparison limitations

The comparative analysis of machine learning algorithms for cloud resource allocation faces fundamental challenges due to the heterogeneous evaluation environments employed across different studies. This section addresses the reliability concerns inherent in cross-paper comparisons and establishes the methodological framework used to mitigate these limitations while providing meaningful insights about algorithmic performance trends. Table 2 summarizes the evaluation environments, datasets, and experimental configurations across all analyzed algorithms, illustrating the substantial heterogeneity that limits direct cross-algorithm comparison.

### 5.1 Evaluation environment heterogeneity

The analyzed algorithms were evaluated using diverse simulation environments, datasets, and experimental configurations that limit direct quantitative comparison. Deep Reinforcement Learning approaches employed different simulation frameworks: PPO for D2D-MEC used custom network simulators for mobile edge computing scenarios, ATSIA3C utilized CloudSim toolkit for multi-cloud environments, and Rainbow DQN implemented specialized IoT device-edge-cloud simulators. Neural Network Architecture methods relied on varied datasets: VSBG used Google cluster traces (672,003 jobs over 29 days) and Alibaba cluster traces (4M jobs over 8 days), DWT-BiGRU employed Google cluster data (672,074 jobs) and Alibaba cluster data (1300 + machines over 12 h), while DPSO-GA utilized Google cluster workload traces with CloudSim simulation.

Traditional Machine Learning Enhanced approaches employed domain-specific evaluation environments: Enhanced-Kernel SVM used iFogSim simulator for healthcare fog-cloud scenarios with 20–60 sensors, while N2TC-GATA utilized Google cluster-traces v3 dataset with 405,894 task records. Multi-Agent Based approaches required specialized distributed simulation: Multi-Agent DRL employed real private cloud environments with 8 heterogeneous servers, while IF-DDPG used industrial IoT simulation with four edge servers and four terminal devices across $100 \times 100$ m$^2$ factory areas.

The analyzed algorithms were evaluated across fundamentally different application environments that create distinct resource demand patterns, significantly limiting cross-algorithm comparison validity. Healthcare applications (Enhanced-Kernel SVM) exhibit ultra-low latency requirements and predictable monitoring patterns, mobile environments (PPO, Rainbow DQN) demonstrate high mobility with variable connectivity, industrial IoT (IF-DDPG) requires deterministic real-time processing, and enterprise environments (ATSIA3C, N2TC-GATA) show complex workload patterns with cost optimization priorities. These domain-specific characteristics mean algorithms optimized for healthcare temporal patterns may perform poorly in enterprise bulk processing scenarios, and mobile energy-constrained optimizations may not apply to industrial fault-tolerant requirements.

TABLE 2 Evaluation environments, datasets, and experimental configurations for analyzed algorithms.

| Algorithm | Evaluation environment | Dataset/traces | Scale | Baseline algorithms | Key metrics |
|---|---|---|---|---|---|
| PPO for D2D-MEC | Custom mobile edge computing simulator | Synthetic mobile device traces | 100 + mobile devices, 10 edge servers | Random offloading, local execution, always-cloud | Execution time, energy consumption, monetary cost |
| ATSIA3C | CloudSim toolkit | Synthetic multi-cloud workloads | 1000 + tasks across multiple datacenters | RATS-HM, MOABCQ, AINN-BPSO | Makespan, resource cost, energy consumption |
| Rainbow DQN | Custom IoT-edge-cloud simulator | Synthetic IoT device traces | 100–300 IoT devices, 3-tier architecture | Double DQN, standard DQN | Energy efficiency, latency, utility |
| DPSO-GA | CloudSim simulation | Google cluster workload traces | 50–1500 tasks, 10–50 VMs | Standalone PSO, standalone GA | Waiting time, energy consumption, prediction MAE |
| VSBG | Custom prediction framework | Google cluster traces (672,003 jobs, 29 days), Alibaba cluster traces (4M jobs, 8 days) | Large-scale cluster workloads | ARIMA, standard LSTM | RMSLE, MSE, $R^2$ coefficient |
| DWT-BiGRU | Custom prediction framework | Google cluster (672,074 jobs, 29 days), Alibaba cluster (1300 + machines, 12 h) | Multi-machine cluster environments | SVR, standard LSTM | MAPE, RMSE |
| Enhanced-Kernel SVM | iFogSim simulator | Synthetic e-health application traces | 20–60 sensors, fog-cloud hybrid | FCFS, cloud-only architecture | Execution time, latency |
| N2TC-GATA | Custom scheduling simulator | Google cluster-traces v3 (405,894 tasks) | Large-scale task scheduling | Traditional genetic algorithm | Execution time, cost efficiency, response time |
| Multi-agent DRL | Real private cloud environment | Real container workloads | Eight heterogeneous servers, container batches | Kubernetes, Best-Fit, Max-Fit | Overall runtime, container placement efficiency |
| IF-DDPG | Custom industrial IoT simulator | Synthetic industrial workloads | Four edge servers, four terminal devices, $100 \times 100$ m$^2$ area | Traditional DDPG, DDQN, DQN, actor-critic | Energy consumption |

Organizations should prioritize algorithms evaluated in application environments similar to their target deployment context rather than relying on performance improvements reported across different domains.

## 5.2 Baseline algorithm variability

The performance improvements reported across studies derive from comparisons against different baseline algorithms, creating additional complexity in cross-study evaluation. Deep Reinforcement Learning algorithms compared against varied baselines: ATSIA3C benchmarked against RATS-HM, MOABCQ, and AINN-BPSO algorithms, while Rainbow DQN used Double DQN as primary comparison. Neural Network approaches employed different baseline sets: VSBG compared against ARIMA and standard LSTM models, DWT-BiGRU benchmarked against SVR and basic LSTM implementations, and DPSO-GA evaluated against standalone PSO and GA algorithms.

Traditional ML Enhanced methods used domain-specific baselines: Enhanced-Kernel SVM compared against FCFS algorithms and cloud-only architectures, while N2TC-GATA benchmarked against traditional genetic algorithms without neural network classification. Multi-Agent approaches required specialized comparison frameworks: Multi-Agent DRL compared against Kubernetes, Best-Fit, and Max-Fit algorithms, while IF-DDPG benchmarked against traditional DDPG, DDQN, DQN, and Actor-Critic algorithms.

## 5.3 Performance metric standardization challenges

The diversity of performance metrics across studies reflects different optimization objectives and evaluation priorities, complicating direct algorithmic comparison. Execution time and makespan reduction metrics vary significantly in measurement approaches: some studies report absolute time improvements while others focus on percentage reductions relative to baseline algorithms. Cost optimization metrics employ different economic models: ATSIA3C measures resource cost optimization across multi-cloud pricing models, while Enhanced-Kernel SVM focuses on operational cost reduction in healthcare environments.

Energy efficiency measurements demonstrate substantial methodological differences: IF-DDPG reports energy consumption reduction in industrial IoT scenarios measured in watts, ATSIA3C calculates energy improvements as percentage reductions across datacenter operations, while DWT-BiGRU focuses on computational energy efficiency during prediction tasks. Quality of Service metrics vary by application domain: healthcare applications emphasize latency reduction and reliability measures, while industrial IoT focuses on real-time processing capabilities and fault tolerance.

The focus on best performance metrics for each algorithm reflects the inherent challenge of comparing algorithms optimized for different objectives and evaluated using heterogeneous methodologies. Each algorithm prioritizes specific optimization goals: ATSIA3C emphasizes makespan reduction in multi-cloud environments, VSBG targets prediction accuracy for workload forecasting, Enhanced-SVM optimizes execution time for healthcare applications, and IF-DDPG focuses on energy efficiency in industrial settings. Rather than attempting misleading cross-metric comparisons between fundamentally different optimization objectives, our analysis emphasizes each algorithm's primary contribution within its intended application domain. This approach provides more meaningful insights about algorithmic effectiveness while acknowledging that comprehensive multi-metric comparison would require standardized evaluation environments and consistent optimization objectives across all studies—conditions that do not exist in the current literature.

## 5.4 Mitigation strategies and analytical framework

To address these comparison limitations, our analysis employs several mitigation strategies that enable meaningful insights while acknowledging methodological constraints. Relative performance analysis focuses on percentage improvements over baseline algorithms reported in original studies rather than attempting direct comparison of absolute performance values. This approach enables identification of algorithmic effectiveness trends while respecting the different evaluation contexts employed across studies.

Categorical performance grouping organizes algorithms by problem type and application domain, enabling more appropriate comparisons within similar operational contexts. Deep Reinforcement Learning algorithms are analyzed collectively for their learning efficiency and adaptation capabilities, Neural Network approaches are compared for prediction accuracy and temporal modeling effectiveness, Traditional ML Enhanced methods are evaluated for domain-specific optimization performance, and Multi-Agent approaches are assessed for distributed coordination effectiveness.

Qualitative capability assessment examines algorithmic features and architectural innovations rather than relying solely on quantitative performance metrics. This includes analysis of convergence properties, scalability characteristics, implementation complexity, and deployment requirements that affect practical adoption decisions. The assessment framework considers algorithmic robustness, adaptation capabilities, and integration requirements that influence real-world effectiveness beyond simulation-based performance metrics.

## 5.5 Reliability assessment and confidence levels

Given the evaluation methodology limitations, performance comparisons should be interpreted with appropriate confidence levels and uncertainty acknowledgment. High confidence comparisons are possible within algorithm categories where studies employ similar evaluation environments and baseline algorithms, such as Neural Network approaches using Google cluster datasets or Deep Reinforcement Learning methods evaluated in mobile edge computing scenarios.

Medium confidence comparisons apply across categories where different evaluation environments prevent direct quantitative comparison but similar problem formulations enable qualitative capability assessment. This includes comparisons between optimization objectives such as energy efficiency improvements across different algorithmic approaches or scalability characteristics between centralized and distributed optimization methods.

Low confidence comparisons acknowledge situations where evaluation environments differ significantly, preventing reliable performance assessment. Cross-domain comparisons between healthcare-specific algorithms and general-purpose resource allocation methods fall into this category, as do comparisons between simulation-based results and real-world deployment outcomes.

## 5.6 Implications for algorithmic selection

The evaluation reliability analysis provides important guidance for algorithmic selection and implementation decisions. Organizations should prioritize algorithms evaluated in environments similar to their deployment contexts rather than relying solely on reported performance improvements. The heterogeneous evaluation landscape emphasizes the importance of pilot testing and performance validation in target environments before full-scale deployment.

Algorithm selection should consider evaluation methodology quality alongside reported performance metrics, with preference for approaches validated across multiple datasets and evaluation environments. The analysis suggests that algorithmic robustness and adaptation capabilities may be more important than peak performance metrics when deployment environments differ from evaluation contexts used in original studies.

# 6 Cross-category comparative analysis

This section presents a comparative analysis across all four categories of machine learning algorithms for cloud resource allocation, examining performance characteristics, implementation considerations, and practical deployment guidance. Through systematic evaluation of the 10 algorithms, we identify key trends, architectural patterns, and optimization strategies that characterize the current state-of-the-art in intelligent cloud resource management.

## 6.1 Performance trade-offs and the performance-complexity paradox

The comparative analysis reveals fundamental trade-offs between algorithmic categories that directly impact their suitability for different cloud environments, uncovering a critical performance-complexity paradox where algorithmic sophistication often correlates inversely with practical deployability. Deep Reinforcement Learning algorithms demonstrate the most substantial performance gains across multiple metrics, with ATSIA3C achieving remarkable 70.49% makespan reduction, 77.42% cost optimization, and 74.24% energy consumption improvement. However, these impressive results exemplify the performance-complexity paradox: improvements exceeding 50% typically require infrastructure investments and operational expertise that may not justify incremental benefits for many deployment scenarios.

The paradox manifests through three distinct dimensions affecting adoption decisions. Implementation complexity increases exponentially with performance gains, as ATSIA3C's superior results require mastery of residual CNN architectures, A3C algorithm implementation, and multi-cloud simulation environments. Operational overhead scales non-linearly with algorithmic sophistication, with maintenance, retraining, and continuous optimization demanding specialized expertise often unavailable in typical IT operations teams. Infrastructure requirements grow substantially with performance expectations, as advanced algorithms require specialized hardware, distributed training capabilities, and extensive computational resources that may exceed organizational capacity.

This analysis reveals that practical algorithmic selection should prioritize robustness and deployability over peak performance metrics when operational constraints limit implementation sophistication. The sample efficiency problem remains particularly acute, with PPO requiring 500–1,000 episodes for convergence, making real-time adaptation challenging in dynamic cloud environments while demanding continuous computational resources for effective operation.

Figure 2 illustrates the positioning of analyzed algorithms across two critical dimensions for practical deployment: performance impact and implementation complexity. This qualitative assessment reveals four distinct clusters that inform algorithmic selection strategies based on organizational constraints and performance requirements.

The quadrant analysis reveals that high-impact algorithms (ATSIA3C, VSBG, IF-DDPG) generally require substantial implementation complexity, while simpler approaches (Enhanced-SVM, N2TC-GATA) vary significantly in their performance contributions. This positioning supports the performance-complexity paradox identified in our analysis, where breakthrough performance often correlates with implementation challenges that may limit practical adoption.

Neural Network Architecture approaches excel in prediction accuracy but suffer from different limitations. VSBG achieves exceptional workload prediction with RMSLE improvements from 0.89 to 0.03, yet this accuracy relies heavily on historical data patterns that may not generalize to novel workload behaviors or sudden system changes. The signal processing components (VMD, DWT) introduce additional computational complexity and parameter sensitivity that can degrade performance when applied to workloads with different characteristics than training data. Furthermore, the temporal dependencies captured by LSTM and BiGRU architectures assume stationarity in underlying system behavior that may not hold in rapidly evolving cloud environments.

Traditional Machine Learning Enhanced approaches offer more predictable performance with lower computational requirements, but their effectiveness remains highly domain-specific. Enhanced-Kernel SVM achieves 73.88% execution time improvement for healthcare applications, yet the specialized kernel design limits generalizability across different application domains. The hybrid optimization techniques in N2TC-GATA provide balanced multi-objective solutions but struggle with scalability when the number of objectives or constraints increases significantly beyond the evaluated scenarios.

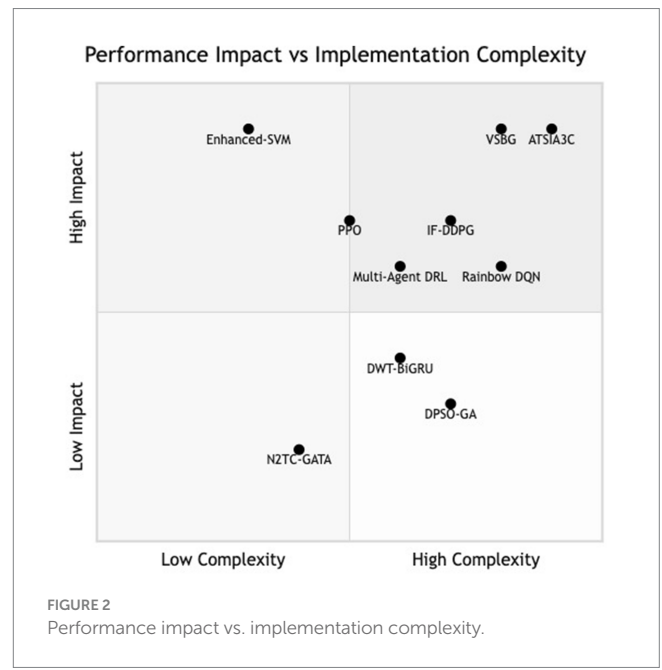

FIGURE 2
Performance impact vs. implementation complexity.

Multi-Agent Based approaches demonstrate superior distributed coordination but introduce complex synchronization challenges. Multi-Agent DRL achieves 28% runtime improvement through intelligent coordination, yet the sequential decision-making mechanism creates potential bottlenecks as the number of agents increases. The federated learning component in IF-DDPG provides privacy preservation but introduces communication overhead and convergence delays that may offset performance gains in high-frequency resource allocation scenarios.

## 6.2 Convergence properties and scalability analysis

A critical examination of convergence properties reveals significant variations in algorithmic stability and scalability characteristics across categories, with fundamental theoretical limitations that constrain practical deployment at enterprise scale. Deep Reinforcement Learning algorithms exhibit non-monotonic convergence behavior with potential for temporary performance degradation during exploration phases, creating operational risks in production environments where consistent performance is essential. ATSIA3C demonstrates faster convergence through residual CNN integration, yet the algorithm's performance remains critically sensitive to reward function design and hyperparameter configuration, suggesting fundamental brittleness that limits reliable deployment.

The scalability analysis reveals architectural limitations that represent theoretical rather than implementation barriers. Deep Reinforcement Learning algorithms face quadratic complexity growth with state space expansion, creating mathematical constraints that cannot be overcome through hardware improvements or optimization techniques. For state spaces characteristic of enterprise cloud environments with millions of resources and complex interdependencies, current DRL approaches would require computational resources that exceed physically feasible limits. Rainbow DQN's six enhancement techniques improve stability but increase memory requirements exponentially with state space dimensionality, creating a fundamental trade-off between algorithmic sophistication and scalability that cannot be resolved within current architectural paradigms.

Neural Network Architecture approaches generally demonstrate more predictable convergence but suffer from generalizability limitations that restrict real-world applicability. VSBG's multi-modal prediction achieves stable performance across Google and Alibaba datasets, yet the decomposition techniques lose effectiveness when workload patterns deviate significantly from training distributions, revealing fundamental assumptions about data stationarity that rarely hold in dynamic cloud environments. The temporal dependencies captured by LSTM and BiGRU architectures assume underlying system behavior patterns that may not persist across different deployment contexts, geographical regions, or temporal periods, limiting algorithmic robustness to specific evaluation scenarios.

Traditional ML Enhanced methods scale more predictably but encounter fundamental limitations when problem complexity exceeds their theoretical foundations. Enhanced-Kernel SVM's domain-specific optimization relies on kernel design assumptions that cannot generalize beyond healthcare applications without complete algorithmic redesign. The mathematical foundations of SVM classification create inherent boundaries on problem complexity that cannot be exceeded regardless of computational resources or data availability, suggesting that traditional approaches face theoretical rather than practical scalability constraints.

Multi-Agent Based approaches present unique scalability challenges related to coordination complexity rather than computational requirements. Multi-Agent DRL's sequential decision-making mechanism creates coordination bottlenecks that scale exponentially with agent population, fundamentally limiting the approach to scenarios with modest numbers of coordinating entities. The communication overhead in federated learning approaches like IF-DDPG grows quadratically with participant numbers, creating theoretical limits on distributed optimization that cannot be overcome through network improvements or protocol optimization.

Analysis reveals three fundamental challenges that transcend individual algorithmic approaches. The adaptability-stability paradox manifests differently across categories: DRL algorithms achieve adaptability through continuous learning but risk performance degradation during exploration, neural networks provide stable predictions but require complete retraining when patterns change, traditional ML offers predictable performance but lacks adaptation mechanisms, and multi-agent systems demonstrate coordination flexibility but struggle with protocol evolution. The scale-complexity ceiling creates theoretical barriers where DRL faces quadratic state space growth, neural networks suffer from dimensionality curses, traditional ML encounters mathematical complexity limits, and multi-agent coordination overhead grows exponentially. These systematic limitations suggest that breakthrough advances require interdisciplinary approaches combining control theory, distributed systems, and optimization theory rather than incremental improvements within existing paradigms.

## 6.3 Critical gap analysis and systematic field limitations

Despite impressive individual performance improvements, the analyzed algorithms collectively reveal systematic limitations that persist across all categories and represent fundamental challenges requiring theoretical advancement rather than incremental algorithmic refinement. The evaluation methodology inconsistency across studies creates uncertainty about real-world performance, with different baseline comparisons, simulation environments, and performance metrics preventing reliable cross-algorithm evaluation. This evaluation validity crisis undermines confidence in reported improvements and suggests the field requires standardized frameworks before meaningful algorithmic comparison becomes possible.

The temporal adaptation challenge represents a fundamental limitation affecting all categories but manifesting through different mechanisms. Deep reinforcement learning algorithms require extensive retraining when environment dynamics change, neural network approaches suffer from concept drift when workload patterns evolve beyond training distributions, traditional ML methods lack adaptive mechanisms for changing optimization landscapes, and multi-agent systems struggle with coordination protocol adaptation during system evolution. This systematic challenge suggests that current approaches fundamentally misunderstand the dynamic nature of cloud environments, treating them as stationary systems rather than continuously evolving ecosystems.

The scalability ceiling emerges as a critical architectural limitation where current approaches demonstrate effectiveness for hundreds to thousands of resources but lack theoretical or empirical validation for million-scale deployments characteristic of major cloud providers. None of the analyzed algorithms provide convincing evidence of linear scalability, with most exhibiting exponential complexity growth that limits practical applicability to large-scale production environments. This limitation indicates that fundamental algorithmic paradigms may require replacement rather than refinement to address enterprise-scale cloud resource allocation demands.

Resource allocation algorithms also suffer from the multi-objective optimization paradox, where improvements in one metric often come at the expense of others, creating trade-offs that current approaches handle poorly. Cost optimization frequently conflicts with performance maximization, energy efficiency may reduce service quality, and security constraints limit optimization flexibility. The analyzed algorithms typically optimize for specific objective combinations but lack principled frameworks for dynamically adjusting optimization priorities based on changing business requirements or system conditions, suggesting fundamental theoretical gaps in multi-objective resource allocation.

## 6.4 Generalizability crisis and deployment reality gap

The systematic analysis reveals a fundamental generalizability crisis where algorithms demonstrate impressive performance in controlled evaluation environments but face severe limitations when deployed in real-world cloud infrastructures with different characteristics, constraints, and operational requirements. This deployment reality gap manifests through several critical dimensions that question the practical applicability of current research directions.

Dataset dependency limitations plague all analyzed algorithms, with performance claims based on specific historical traces that may not represent future workload patterns or different organizational contexts. VSBG's exceptional performance on Google and Alibaba cluster traces cannot reliably predict effectiveness for financial services workloads, healthcare applications, or e-commerce platforms with fundamentally different usage patterns, temporal characteristics, and resource consumption behaviors. The implicit assumption that historical patterns predict future performance ignores the rapidly evolving nature of cloud applications, particularly with the integration of AI/ML workloads that exhibit fundamentally different resource consumption characteristics than traditional enterprise applications.

Environmental constraint sensitivity represents a critical limitation where algorithms optimized for specific infrastructure configurations fail when deployed in environments with different hardware characteristics, network topologies, or resource availability patterns. Enhanced-Kernel SVM's healthcare optimization assumes fog-cloud architectures with specific latency and bandwidth characteristics that may not exist in rural healthcare deployments, edge computing scenarios with limited connectivity, or regulatory environments requiring data locality constraints. The algorithmic assumptions about available infrastructure create fundamental barriers to cross-domain deployment that cannot be resolved through parameter tuning or minor modifications.

Operational context brittleness emerges as algorithms assume specific operational procedures, maintenance windows, and failure patterns that may not align with actual deployment environments. Multi-Agent DRL's coordination mechanisms assume reliable inter-agent communication that may not hold in distributed cloud environments with network partitions, variable latency, or security constraints that limit information sharing between components. The gap between algorithmic assumptions and operational reality creates deployment risks that may outweigh performance benefits.

Scale transition failures occur when algorithms demonstrate effectiveness at research scales but encounter fundamental barriers when deployed at enterprise scales with millions of resources, complex organizational structures, and regulatory compliance requirements. The mathematical complexity growth patterns identified in scalability analysis suggest that current algorithmic paradigms cannot bridge the gap between research demonstrations and production deployment requirements, necessitating fundamental theoretical advances rather than incremental improvements.

This generalizability crisis indicates that the field may be optimizing for evaluation success rather than practical deployment effectiveness, suggesting a fundamental misalignment between research objectives and industry needs that requires systematic correction through deployment-focused evaluation frameworks and real-world validation requirements.

Systematic analysis reveals that improvements in one performance dimension often conflict with others across all algorithmic categories. Cost optimization frequently conflicts with performance maximization, energy efficiency may reduce service quality, and security constraints limit optimization flexibility. Current algorithms typically optimize for specific objective combinations but lack principled frameworks for dynamically adjusting optimization priorities based on changing business requirements or system conditions, suggesting fundamental theoretical gaps in multi-objective resource allocation that require new mathematical frameworks rather than incremental algorithmic improvements.

## 6.5 Strategic implementation guidelines

The comparative analysis enables the development of strategic guidelines for algorithm selection based on organizational requirements and deployment constraints. Organizations with mature MLOps infrastructure and tolerance for complex deployment procedures should consider Deep Reinforcement Learning approaches for scenarios requiring maximum performance optimization across multiple objectives. The substantial training overhead and infrastructure requirements make these approaches suitable for large-scale deployments where the performance benefits justify the implementation complexity.

Neural Network Architecture approaches provide optimal solutions for organizations with predictable workload patterns and historical data availability. These algorithms excel in environments where accurate demand forecasting directly translates to cost savings and performance improvements. However, organizations must invest in data collection infrastructure and model maintenance procedures to ensure continued effectiveness as system characteristics evolve.

Traditional Machine Learning Enhanced methods offer the most accessible entry point for organizations seeking immediate improvements with existing infrastructure. The interpretable nature of SVM-based approaches and hybrid optimization techniques provides transparency essential for regulatory compliance and operational validation. These approaches work best in specialized domains with well-defined optimization objectives and constraint sets.

Multi-Agent Based approaches suit organizations operating distributed cloud environments with strong privacy requirements or regulatory constraints limiting data sharing. The federated learning capabilities enable collaborative optimization while preserving data sovereignty, making these approaches particularly valuable for multi-organizational cloud federations or industry consortiums requiring coordinated resource management.

## 6.6 Synthesis of hybrid architecture effectiveness and evolutionary principles

The analysis reveals that hybrid architectures consistently outperform single-method approaches across all performance metrics and deployment scenarios, representing not merely incremental improvement but fundamental evolution in algorithmic design philosophy. This effectiveness stems from three systematic integration mechanisms that address the inherent limitations of isolated AI/ML techniques. Capability augmentation combines techniques with non-overlapping strengths, such as DPSO-GA's integration of CNN spatial feature extraction with PSO-GA global optimization, creating synergistic effects that exceed individual component capabilities. Weakness mitigation pairs techniques where one addresses the other's limitations, exemplified by VSBG's combination of VMD signal decomposition with hybrid LSTM architectures to handle nonstationary data that neither component could manage effectively in isolation. The analysis reveals that hybrid architectures consistently outperform single-method approaches across all performance metrics and deployment scenarios, representing not merely incremental improvement but fundamental evolution in algorithmic design philosophy. This effectiveness stems from three systematic integration mechanisms that address the inherent limitations of isolated AI/ML techniques. Capability augmentation combines techniques with non-overlapping strengths, such as DPSO-GA's integration of CNN spatial feature extraction with PSO-GA global optimization, creating synergistic effects that exceed individual component capabilities. Weakness mitigation pairs techniques where one addresses the other's limitations, exemplified by VSBG's combination of VMD signal decomposition with hybrid LSTM architectures to handle nonstationary data that neither component could manage effectively in isolation.

Performance amplification leverages emergent properties where combined techniques achieve superior results through complex interactions rather than simple addition of individual contributions, demonstrated by ATSIA3C's residual CNN integration with A3C that improves both feature extraction and policy learning simultaneously through architectural co-evolution. This mechanism suggests that optimal hybrid design requires understanding technique interactions rather than merely combining high-performing components.

The architectural evolution patterns reveal a clear trajectory from reactive single-technique approaches toward predictive hybrid systems that integrate multiple AI/ML paradigms following identifiable principles. This convergence occurs systematically because modern cloud environments exhibit complexity characteristics that exceed the capability boundaries of individual machine learning paradigms. The fundamental principle emerging from this analysis suggests that optimal cloud resource allocation requires integration of complementary AI/ML capabilities rather than reliance on single techniques, with the most successful architectures achieving balance between capability breadth, implementation complexity, and operational maintainability.

However, hybrid architectures introduce their own systematic challenges including increased implementation complexity, higher computational requirements, and more complex failure modes that can offset performance benefits in resource-constrained environments. The integration overhead creates a critical threshold effect where architectural sophistication must be balanced against deployment constraints, suggesting that successful hybrid design requires principled approaches to complexity management rather than unbounded technique integration. Organizations considering hybrid approaches must carefully evaluate whether performance benefits justify additional operational complexity and resource requirements, particularly when deployment environments impose strict resource or expertise constraints.

The convergence of challenges across all categories illuminates requirements for next-generation systems: (1) temporal adaptation mechanisms that treat dynamic environments as fundamental design constraints, (2) hierarchical architectures achieving linear scalability through principled decomposition, and (3) multi-objective frameworks enabling dynamic priority adjustment based on operational context. Future research should prioritize hybrid architectures integrating multiple ML paradigms, meta-learning approaches for rapid adaptation, and quantum-inspired optimization for addressing exponential complexity challenges.

# 7 Future work

The analysis of machine learning-based cloud resource allocation algorithms reveals significant opportunities for advancing the field through targeted research initiatives. Future research should prioritize next-generation hybrid architectures that systematically integrate multiple machine learning paradigms, meta-learning approaches for rapid adaptation to new cloud environments, and quantum-inspired optimization algorithms for addressing exponential complexity challenges.

Current approaches demonstrate effectiveness for hundreds to thousands of resources, but modern cloud environments require fundamentally different approaches for millions of resources. Research should explore hierarchical multi-agent architectures, enhanced federated learning with secure aggregation protocols and differential privacy mechanisms (Dwork, 2008), and edge-cloud-quantum computing integration for unified optimization frameworks.

Domain-specific applications require specialized algorithms addressing medical data processing and regulatory compliance (Dwork, 2008; NIST, 2020) for healthcare environments, operational technology integration for industrial IoT, and multi-objective optimization for smart city applications. Security research should focus on zero-trust computing environments, sustainability initiatives with carbon-aware algorithms, and 5G/6G network integration for ultra-low latency requirements.

# 8 Conclusion

This comprehensive survey presents a systematic analysis of 10 state-of-the-art machine learning algorithms for cloud resource allocation across four methodological categories: Deep Reinforcement Learning, Neural Network Architecture, Traditional Machine Learning Enhanced, and Multi-Agent Based approaches.

The analysis reveals that machine learning-based approaches consistently outperform traditional heuristic methods, with improvements ranging from 10% to over 70%. Deep Reinforcement Learning algorithms demonstrate the most substantial gains, with ATSIA3C achieving 70.49% makespan reduction, 77.42% cost optimization, and 74.24% energy consumption improvement. Neural Network Architecture methods excel in prediction accuracy, with VSBG achieving RMSLE improvements from 0.89 to 0.03. Traditional ML Enhanced approaches prove effective for domain-specific applications, with Enhanced-Kernel SVM demonstrating 73.88% execution time improvement for healthcare tasks. Multi-Agent approaches show superior distributed coordination, with Multi-Agent DRL achieving 28% runtime improvement and IF-DDPG delivering 15.2 to 50.5% energy reductions.

Key trends include the emergence of hybrid architectures as dominant approaches, edge computing integration across multiple algorithms, and privacy preservation through federated learning capabilities. The practical analysis reveals varying implementation complexity and computational requirements, providing selection guidance for organizations with different deployment constraints and performance objectives.

This survey contributes a comparative analysis of recent ML approaches for cloud resource allocation, offering technical insights for researchers and practical guidance for industry practitioners. The systematic evaluation framework enables informed decision-making for implementing next-generation cloud resource allocation strategies. As cloud computing evolves toward increasingly complex and distributed architectures, the insights presented guide development of intelligent systems capable of meeting modern computing challenges through the fundamental transformation from reactive to predictive and adaptive resource allocation.

## Author contributions

DB: Conceptualization, Data curation, Formal analysis, Investigation, Methodology, Project administration, Resources, Supervision, Validation, Visualization, Writing – original draft, Writing – review & editing. SK: Data curation, Formal analysis, Investigation, Validation, Writing – original draft, Writing – review & editing.

## Funding

The author(s) declare that no financial support was received for the research and/or publication of this article.

## Conflict of interest

The authors declare that the research was conducted in the absence of any commercial or financial relationships that could be construed as a potential conflict of interest.

## Generative AI statement

The authors declare that no Gen AI was used in the creation of this manuscript.

Any alternative text (alt text) provided alongside figures in this article has been generated by Frontiers with the support of artificial intelligence and reasonable efforts have been made to ensure accuracy, including review by the authors wherever possible. If you identify any issues, please contact us.

## Publisher's note